\documentstyle[pra,aps]{revtex}

\newtheorem{theorem}{Theorem}
\newtheorem{corollary}{Corollary}
\newtheorem{lemma}{Lemma}
\newtheorem{proposition}{Proposition}
\newtheorem{definition}{Definition}
\newtheorem{example}{Example}

\begin{document}

\title{ Unextendible product bases and the construction of inseparable
states.}
\author{Arthur O. Pittenger}
\address{Department of Mathematics and Statistics\\
University of Maryland, Baltimore County\\
Baltimore, Maryland 21250}
\maketitle

\begin{abstract}
Let $H^{\left( N\right) }$ denote the tensor product of $n$ finite
dimensional Hilbert spaces $H^{\left( r\right) }$. A state $\left| \varphi
\right\rangle $ of $H^{\left( N\right) }$ is separable if $\left| \varphi
\right\rangle =\left| \alpha _{1}\right\rangle \otimes \cdots \otimes \left|
\alpha _{n}\right\rangle $ where the states $\left| \alpha _{r}\right\rangle 
$ are in $H^{\left( r\right) }$. An \textit{orthogonal} \textit{unextendible
product basis} is a finite set $B$ of separable orthonormal states $\left\{
\left| \varphi _{k}\right\rangle ,1\leq k\leq m\right\} $ such that the
non-empty space $B^{\bot }$, the set of vectors orthogonal to $B$, contains
no separable state. Examples of orthogonal \textit{UPB} sets were first
constructed by Bennett et al \cite{BD1} and other examples and references
appear, for example, in \cite{diVTer}. If $F=F\left( B\right) $ denotes the
set of convex combinations of $\left\{ \left| \varphi _{k}\right\rangle
\left\langle \varphi _{k}\right| ,1\leq k\leq m\right\} $, then $F$ is a
face in the set $S$ of separable densities. In this note we show how to use $%
F$ to construct families of positive partial transform states (\textit{PPT})
which are not separable. We also show how to make an analogous construction
when the condition of orthogonality is dropped. The analysis is motivated by
the geometry of the faces of the separable states and leads to a natural
construction of entanglement witnesses separating the inseparable \textit{PPT%
} states from \textit{S}.
\end{abstract}

\section{\textbf{Background}}

The basic mathematical context of quantum computing and quantum information
theory involves a tensor product of Hilbert spaces 
\[
H^{\left[ N\right] }=H^{\left[ d_{1}\right] }\otimes \ldots \otimes
H^{\left[ d_{n}\right] }, 
\]
and one of the operational aspects of the theory is the feature of \textit{%
entanglement} of different factors of the tensor product. The mathematical
expression of this feature involves the subset of $D$, the set of trace one
positive semidefinite operators or densities on $H^{\left[ N\right] }$,
which are not in $S$, the subset of \textit{separable} densities defined as
the convex hull of rank one separable projections of the form 
\[
P=\left| \varphi \right\rangle \left\langle \varphi \right| =\otimes
_{k=1}^{n}\left| \alpha _{k}\right\rangle \left\langle \alpha _{k}\right| . 
\]
In this notation, $P$ denotes the projection as an operator on $H^{\left[
N\right] }$, $\left| \varphi \right\rangle $ denotes in Dirac notation a
normalized non-null eigenvector with $\left| \varphi \right\rangle
\left\langle \varphi \right| $ the Dirac outer product notation for a rank $%
1 $ projection, and separability appears in the requirement that $\left|
\varphi \right\rangle $ is the tensor product of vectors $\left| \alpha
_{k}\right\rangle $ in $H^{\left[ d_{k}\right] }$. The problem of
determining whether a density $\rho $ in the compact, convex set $D$ is also
in the smaller compact, convex set $S$ is known as the ``separability
problem'' and has been the subject of much recent research in the quantum
computing literature.

All of this abstraction conceals the very real technical problem of
constructing in the laboratory a physical entity whose representation is an
inseparable density $\rho $ and which has the potential of experimentally
realizing some of the rather bizarre predictions of quantum mechanics. In
particular, in some circumstances the resulting entanglement between two
distinct physical systems can be used as a resource to demonstrate
``non-local'' behavior between the two systems which may be physically quite
far apart. In practice, that means that measurements of the two distinct
systems are correlated in ways which cannot be explained by an
interpretation based on classical theory.

In 1994 Peter Shor \cite{shor1} defined an algorithm which could use the
quantum mechanical properties of superposition and entanglement to determine
the prime factors of a large number $M$. Since the work factor of the
algorithm was polynomial in the number of digits of $M$, a significant
improvement over the best classical factoring algorithms, there was
immediate interest in the feasibility of a \textit{quantum computer}, a
computing device able to realize quantum mechanical entanglement. As a
result, there has been an explosion of theoretical work on the role of
quantum mechanics in areas such as computing, cryptography, information
theory and complexity theory, and a corresponding growth of experimental
work directed at demonstrating some of the theoretical predictions.

In this paper we concentrate exclusively on aspects of the separability
problem, and rather than try to summarize all of the relevant references to
that subject and to the motivating work mentioned above, we refer the reader
to \cite{NC} and \cite{pre1} for references and a development of all aspects
of the theory and to \cite{pitt1} which concentrates on the development of
quantum computing algorithms and the basics of quantum coding theory. For
references to work on the separability problem we recommend the survey paper 
\cite{ter1} which gives a good overview of the subject.

\section{\textbf{Context}}

Since $H^{\left[ N\right] }$ is finite dimensional, a density $\rho $ is in $%
S$ if and only if $\rho $ can be represented as a finite sum 
\begin{equation}
\rho =\sum_{a}p\left( a\right) \otimes _{k=1}^{n}\left| \alpha _{k}\left(
a\right) \right\rangle \left\langle \alpha _{k}\left( a\right) \right| ,
\label{sepdef}
\end{equation}
using the notation above, where $\sum_{a}p\left( a\right) =1$ with $%
0<p\left( a\right) \leq 1$. In \cite{peres} Peres observed that a necessary
condition for $\rho $ to be separable is that its \textit{partial
transpositions} are densities. For a general density (with $n=2$), if one
writes $\rho $ as a matrix in a coordinate basis and indexing which respects
the tensor product, then the $\left( i_{1}i_{2},j_{1}j_{2}\right) $'th entry
of the partial transpose $\rho ^{T_{2}}$ is the $\left(
i_{1}j_{2},j_{1}i_{2}\right) $'th entry of $\rho $. For a separable density
one can use complex conjugation and the Hermicity of $\rho $ to write the
partial transposition as 
\[
\rho ^{T_{2}}=\sum_{a}p\left( a\right) \left| \alpha _{1}\left( a\right)
\right\rangle \left\langle \alpha _{1}\left( a\right) \right| \otimes \left|
\alpha _{2}^{*}\left( a\right) \right\rangle \left\langle \alpha
_{2}^{*}\left( a\right) \right| , 
\]
and $\rho ^{T_{2}}$ is also a density. In the general case, the superscript
will denote partial transposition with respect to a subset of the indices,
and the generalized Peres condition is that $\rho ^{T}$ is a density for any
such transposition.

As it happens, in the bivariate $2\otimes 2$ and $2\otimes 3$ cases the
Peres condition is also sufficient: $\rho $ is separable if and only if $%
\rho ^{T_{2}}$ is a density (\cite{HHH1}). However, for all other cases this
is not true: there exist densities which satisfy the Peres condition but
which are not in $S$. Such densities are designated as inseparable $PPT$
(positive partial transform) densities, and it can be shown that physical
systems with these densities do not have the kind of entanglement requisite
for certain kinds of quantum communication \cite{H2}. (See \cite{ter1} for
an exposition and references.)

It is obviously of interest to be able to characterize such $PPT$ densities,
and, correspondingly, it is useful to have a way of explicitly constructing
examples. An important source of examples is based on the idea of an
orthogonal\textit{\ unextendible product basis} \cite{BD1}. (Our terminology
differs slightly from that in the existing literature by adding
orthogonality as a separate property.)

\begin{definition}
A set $B$ of separable states 
\[
\left\{ \left| \varphi _{k}\right\rangle =\otimes _{j=1}^{n}\left| \alpha
_{j}\left( k\right) \right\rangle ,1\leq k\leq m\right\} 
\]
is an \textit{unextendible} \textit{product basis} (\textit{UPB}) if the
non-empty space $B^{\bot }$, the set of states orthogonal to all of the $%
\left| \varphi _{k}\right\rangle $, contains no separable state. An \textit{%
orthogonal} \textit{UPB} has the additional constraint that the $\left|
\varphi _{k}\right\rangle $'s are orthogonal.
\end{definition}

In words, this means that one cannot extend the partial basis $B$ by adding
another separable state which is also orthogonal to the states in $B$. At
first glance the construction of such an orthogonal $B$ looks like a
difficult problem, but in \cite{BD1} specific examples are given, and the
methodology was extended by DiVincenzo, Terhal and others. (See \cite{ter1}
for references and \cite{BD1}, \cite{diVTer} and \cite{DiVMor} for examples.)

The relevance of an orthogonal \textit{UPB} is that it is then easy to
construct a specific example of an inseparable density $\rho $ satisfying
the Peres condition \cite{ter2}. Moreover, Terhal also shows that one can
use $\rho $ to construct examples of positive but not completely positive
operators on $B\left( H^{d}\right) $, the set of bounded operators on the
Hilbert space $H^{d}$. We will not go into the definitions and significance
of such positive operators here; suffice it to say that they arise in the
context of C* algebras and were used in \cite{HHH1} to prove the sufficiency
of the Peres condition in the $2\otimes 2$ case.

One way to analyze the densities in $D$ and $S$ is in the context of the
real Hilbert space $M$ which is defined as the set of Hermitian matrices on $%
H^{\left[ N\right] }$ with the trace inner product 
\begin{equation}
\left\langle \left\langle A,B\right\rangle \right\rangle =Tr\left(
A^{\dagger }B\right)  \label{inpord}
\end{equation}
and Hilbert-Schmidt norm $\left\| A-B\right\| =\sqrt{Tr\left( \left(
A-B\right) ^{2}\right) }.$ This approach was taken in \cite{pitrub1} to get
a better perspective of the (Euclidean) geometry of $M$ and the structure of 
$D$ and $S$ in that context. In fact those tools provide a methodology for
finding the nearest separable state to a given inseparable density $\rho $
in particular cases. They also give a way of constructing so-called \textit{%
entanglement} \textit{witnesses,} which are simply Hermitian matrices $W$
defining hyperplanes separating an inseparable $\rho $ from $S$%
\begin{equation}
Tr\left( \rho W\right) <0\leq Tr\left( \sigma W\right) ,all\text{ }\sigma
\in S  \label{entwit}
\end{equation}
with the hyperplane defined as $\left\{ A\in M:Tr\left( AW\right) =0\right\} 
$.

One germane result from \cite{pitrub1} is that if $\tau _{0}$ is the nearest
separable state to a non-separable $\rho _{0}$, then 
\begin{equation}
W_{0}\equiv \tau _{0}+c_{0}I-\rho _{0}  \label{tauwit}
\end{equation}
with $c_{0}=Tr\left( \tau _{0}\left( \rho _{0}-\tau _{0}\right) \right) $ is
an entanglement witness for $\rho _{0}$ and is related to the Euclidean
structure by 
\begin{equation}
Tr\left( \sigma W_{0}\right) =-\left\langle \left\langle \left( \rho
_{0}-\tau _{0}\right) ,\left( \sigma -\tau _{0}\right) \right\rangle
\right\rangle .  \label{witinprod}
\end{equation}
In particular, the separating hyperplane defined by $W_{0}$ contains a face
of $S$: 
\[
F\left( \tau _{0}\right) =\left\{ \sigma :Tr\left( \sigma W_{0}\right)
=0,\sigma \in S\right\} . 
\]

The results in this paper were motivated by combining the techniques and
insights in \cite{ter2} and in \cite{pitrub1}. Specifically we examine the
geometry implicit in Terhal's construction and use the ideas underlying (\ref
{tauwit}) to define a ``geometric'' entanglement witness. We then show how
to construct a collection of inseparable $PPT$ densities near $\rho _{0}$,
again motivated by the geometry, and give a sufficient condition for the
separating hyperplane defined by $\rho _{0}$ to also separate these other $%
PPT$ densities. Using the resulting insights, we can see the consequences of
orthogonality and can give sufficient conditions for comparable
constructions when the hypothesis of orthogonality is dropped. In
particular, these results provide new perspective on the role of faces of $S$
in the analysis of $PPT$ densities.

\section{\textbf{The orthogonal UPB case}}

As above, $B$ denotes an orthogonal unextendible product basis consisting of 
$m$ separable, orthonormal vectors $\left| \varphi _{k}\right\rangle $, and
we define $F\left( B\right) \subset S$ to be the convex hull of the
corresponding projections $\mu _{k}=\left| \varphi _{k}\right\rangle
\left\langle \varphi _{k}\right| $: 
\begin{equation}
F\left( B\right) =\left\{ \mu =\sum_{k=1}^{m}p_{k}\mu
_{k},\sum_{k}p_{k}=1\right\} .  \label{farface}
\end{equation}
A key feature of a density in $F\left( B\right) $ is that its convex
representation is unique and corresponds to its spectral representation. In
fact, $F\left( B\right) $ is a simplex since it is easy to check that each
density in $F\left( B\right) $ has a unique convex representation in terms
of the $\mu _{k}$'s. Letting $D_{0}$ denote the normalized identity $\frac{1%
}{N}I$, define 
\[
\mu _{0}=\sum \frac{1}{m}\mu _{k}\;and\;\rho _{0}=\frac{1}{N-m}\left(
ND_{0}-m\mu _{0}\right) . 
\]

As a first result, we prove that $\rho _{0}$ is an inseparable $PPT$
density, as was shown in \cite{BD1}.

\begin{lemma}
$\rho _{0}$ is an inseparable $PPT$ density on the boundary of $D$.
\end{lemma}

\textit{Proof}: From the orthonormality of the $\left| \varphi
_{k}\right\rangle $'s, 
\[
\left\langle v\right| \rho _{0}\left| v\right\rangle =\frac{1}{N-m}\left(
\langle v\left| v\right\rangle -\sum_{k}\left\langle v\right| \mu _{k}\left|
v\right\rangle \right) \geq 0 
\]
so that $\rho _{0}$ is a density. Since each $\left| \varphi
_{k}\right\rangle $ is in the null space of $\rho _{0}$, $\rho _{0}$ is on
the boundary of $D.$ (See \cite{pitrub1} for the proof that a density is on
the boundary of $D$ if and only if it has a non-trivial null space.) Since
the $\left| \varphi _{k}\right\rangle $'s are separable projections, it is
easy to see that the set $\left\{ \mu _{k}^{T},1\leq k\leq m\right\} $ of
partial transpositions also comes from an unextendible product basis and so
each $\rho _{0}^{T}$ is also a density. Unwinding the notation as in \cite
{ter2}, we see that \medskip $\rho _{0}$ is proportional to the projector on 
$B^{\bot }$, and thus its convex representation cannot include separable
projections. It follows that $\rho _{0}$ is inseparable. (We will give an
alternate proof of inseparability below.)\ $\Box $

We next record a key geometric feature of this setup.

\begin{lemma}
The ``line segment'' from $\mu _{0}$ through $D_{0}$ to $\rho _{0}$ is
orthogonal to $F\left( B\right) $.
\end{lemma}

\textit{Proof}: $D_{0}$ is a convex combination of $\mu _{0}$ and $\rho _{0}$%
, and thus the three are collinear. For each $r$ 
\[
\left\langle \left\langle \left( D_{0}-\mu _{0}\right) ,\left( \mu _{r}-\mu
_{0}\right) \right\rangle \right\rangle =Tr\left( \mu _{0}^{2}\right)
-Tr\left( \mu _{r}\mu _{0}\right) =\frac{1}{m}-\frac{1}{m}=0, 
\]
and by linearity the same is true for all $\sigma $ in $F\left( B\right) $,
completing the proof.\medskip \ $\Box $

The ideas in the next result come from Terhal's work, and the proof uses the
compactness of the set of separable normalized vectors in $H^{\left[
N\right] }$.

\begin{proposition}
inf$\left\{ Tr\left( \mu _{0}\sigma \right) ,\sigma \in S\right\} \equiv 
\frac{\epsilon }{m}>0$ and the non-empty compact, convex subset of $S$%
\[
G\left( B\right) \equiv \left\{ \sigma \in S:Tr\left( \mu _{0}\sigma \right)
=\frac{\epsilon }{m}\right\} 
\]
is contained in an affine set orthogonal to the line segment from $\mu _{0}$
to $\rho _{0}$.
\end{proposition}

\textit{Proof}: By convexity, it suffices to take the infimum over the set
of separable projections. Suppose that infimum were zero. Then there would
be a sequence of separable projections $\left| \psi _{n}\right\rangle $ such
that 
\[
Tr\left( \mu _{0}\left| \psi _{n}\right\rangle \left\langle \psi _{n}\right|
\right) =\frac{1}{m}\sum_{k}\left| \left\langle \varphi _{k}\right| \psi
_{n}\rangle \right| ^{2}\rightarrow 0, 
\]
and by compactness there must be a separable unit vector orthogonal to each
of the $\left| \varphi _{k}\right\rangle $. That contradicts the assumption
of unextendibility, so the infimum is strictly positive and again by
compactness $G\left( B\right) $ must be non-empty. It remains to show the
orthogonality. Let $\sigma _{1}$ and $\sigma _{2}$ be trace one Hermitian
matrices such that $Tr\left( \mu _{0}\sigma _{k}\right) =\frac{\epsilon }{m}$%
. Then 
\[
\left\langle \left\langle \left( \sigma _{1}-\sigma _{2}\right) ,\left( \mu
_{0}-D_{0}\right) \right\rangle \right\rangle =Tr\left( \mu _{0}\sigma
_{1}\right) -Tr\left( \mu _{0}\sigma _{2}\right) =0, 
\]
\medskip completing the proof. \medskip \ $\Box $

There are other geometric aspects of $G\left( B\right) $. For one thing, in
some high-dimensional sense $F\left( B\right) $ and $G\left( B\right) $ are
parallel since they are perpendicular to the one-dimensional affine space
containing $\rho _{0}$, $D_{0}$, and $\mu _{0}$. Also, since for any density 
$\sigma $%
\begin{equation}
\left\langle \left\langle \left( \sigma -D_{0}\right) ,\left( \mu
_{0}-D_{0}\right) \right\rangle \right\rangle =Tr\left( \sigma \mu
_{0}\right) -\frac{1}{N},  \label{traceqn}
\end{equation}
we can interpret the inner product to be that between the two ``vectors'' $%
\sigma -D_{0}$ and $\mu _{0}-D_{0}$ in $M$, and thus $G\left( B\right) $
consists of those separable densities such that 
\[
\left\| \mu _{0}-D_{0}\right\| \left[ \left\| \sigma -D_{0}\right\| \cos
\left( \left( \sigma -D_{0}\right) ,\left( \mu _{0}-D_{0}\right) \right)
\right] =\frac{\epsilon }{m}-\frac{1}{N}
\]
is minimal. Now it is known from a variety of papers, initially in \cite{zuk}
with references and another proof in \cite{pitrub2} and \cite{pitrub1}, that
there is a $D$-neighborhood of the normalized identity $D_{0}$ which is
composed entirely of separable densities. Hence along the line segment from $%
\mu _{0}$ through $D_{0}$ to $\rho _{0}$, there will be a last separable
density $\tilde{\tau}_{0}$ beyond $D_{0}$ and closest to $\rho _{0}$. Thus 
\begin{equation}
\left\langle \left\langle \left( \tilde{\tau}_{0}-D_{0}\right) ,\left( \mu
_{0}-D_{0}\right) \right\rangle \right\rangle <0,  \label{neg}
\end{equation}
implying 
\begin{equation}
0<\frac{N\epsilon }{m}<1.  \label{epsineq}
\end{equation}
Putting this all together we see that $G\left( B\right) $ consists of the
separable densities $\sigma $ which, in terms of their projection on the $%
\rho _{0}$ - $\mu _{0}$ segment, are in the ``farthest'' face from $\mu _{0}$%
.

In defining the entanglement witness in (\ref{tauwit}), one takes the
nearest separable density $\tau _{0}$ as given and then shows the separating
hyperplane contains the analogue of $G\left( B\right) $. In the present
context we already know what the separating hyperplane looks like and define
the analogue of $\tau _{0}$. Specifically set $\tau _{0}\left( s_{0}\right)
=\left( 1-s_{0}\right) D_{0}+s_{0}\rho _{0}$, where $0<s_{0}<1$ is chosen so
that $Tr\left( \tau _{0}\left( s_{0}\right) \mu _{0}\right) =\frac{\epsilon 
}{m}$. Note that we do not claim that $\tau _{0}\left( s_{0}\right) $ itself
is separable.

\begin{proposition}
If $s_{0}=1-\frac{\epsilon N}{m}$, $\tau _{0}=\tau _{0}\left( s_{0}\right) $
and as usual $c_{0}=Tr\left( \tau _{0}\left( \rho _{0}-\tau _{0}\right)
\right) $, then $W_{0}=\tau _{0}+c_{0}I-\rho _{0}$ is an entanglement
witness for $\rho _{0}$.
\end{proposition}

\textit{Proof}: Since $\tau _{0}\left( s_{0}\right) =D_{0}\left( 1+\frac{%
s_{0}m}{N-m}\right) -\frac{s_{0}m}{N-m}\mu _{0}$, one can compute $Tr\left(
\tau _{0}\left( s_{0}\right) \mu _{0}\right) =\frac{\epsilon }{m}$ and 
\[
\frac{1}{N}\left( 1+\frac{s_{0}m}{N-m}\right) -\frac{s_{0}}{N-m}=\frac{%
\epsilon }{m},
\]
giving $s_{0}=1-\frac{\epsilon N}{m}$. Note that $0<s_{0}<1$ follows from (%
\ref{epsineq}). Since 
\[
\rho _{0}-\tau _{0}=\frac{N\epsilon }{N-m}\left( D_{0}-\mu _{0}\right) ,
\]
for separable densities $\sigma $%
\begin{eqnarray}
Tr\left( W_{0}\sigma \right)  &=&-Tr\left[ \left( \rho _{0}-\tau _{0}\right)
\left( \sigma -\tau _{0}\right) \right]   \nonumber \\
&=&\frac{N\epsilon }{N-m}\left[ Tr\left( \mu _{0}\sigma \right) -\frac{%
\epsilon }{m}\right] \geq 0.  \label{sepcon}
\end{eqnarray}
Since $Tr\left( W_{0}\rho _{0}\right) <0$, the proof is complete. \medskip \ 
$\Box $

The preceding proposition confirms what we already knew - that $\rho _{0}$
is not separable. In later generalizations we will use this approach to
prove inseparability. Before doing that however, let us note that the
geometry also suggests a way of constructing other inseparable $PPT$
densities in the vicinity of $\rho _{0}$. Pictorially, we work with a given $%
\mu \left( p\right) $ in $F\left( B\right) $ and ``reflect'' through $D_{0}$
to obtain a corresponding set of $\rho \left( p\right) $'s including $\rho
_{0}$ on the boundary of $D$. These $\rho \left( p\right) $'s all have
positive partial transforms, and for $\mu \left( p\right) $ in a suitably
small neighborhood of $\mu _{0}$ relative to $F\left( B\right) $ the induced 
$\rho \left( p\right) $'s are also inseparable.

Keeping the same notation, a density in $F\left( B\right) $ can be written
as : 
\begin{equation}
\mu \left( p\right) \equiv \sum_{k}p_{k}\mu _{k}  \label{taup}
\end{equation}
where the $p_{k}$'s are non-negative real numbers with $\sum_{k}p_{k}=1.$
Define $b=b\left( p\right) \equiv 1/\max \left( p_{k}\right) =1/p_{max}$ and 
\begin{equation}
\rho \left( p\right) =\frac{1}{N-b}\left( ND_{0}-b\mu \left( p\right)
\right) .  \label{bdef}
\end{equation}
Note that $b\leq m$ with equality if and only if all of the $p_{k}$'s equal $%
1/m$.

\begin{proposition}
$\rho \left( p\right) $ is a density on the boundary of $D$. If 
\[
p_{max}<\frac{1}{m}+\frac{\epsilon }{m}\left( \frac{N-m}{m-N\epsilon }%
\right) 
\]
then $\rho \left( p\right) $ is an inseparable $PPT$ density .
\end{proposition}

\textit{Proof}: The proof that $\rho \left( p\right) $ is a density with
positive partial transforms is similar to the proof in the first lemma.
Since $\rho \left( p\right) $ has a nontrivial null space containing $\left|
\varphi _{max}\right\rangle $, it's on the boundary of $D$. Finally, from (%
\ref{sepcon}) $Tr\left( \mu _{0}\rho \left( p\right) \right) <\frac{\epsilon 
}{m}$ if and only if $p_{max}$ satisfies the given condition and that gives
inseparability. $\;\Box $

We can put all of these results together to obtain a very nice geometric
result: inseparable $PPT$ states comprise the entire frustram of the cone
with vertex at $D_{0}$, ``base'' defined by the $\rho \left( b\right) $ on
the boundary of $D$ and with the other cross-section defined by the
separating hyperplane defined by $W_{0}$.

\begin{theorem}
If $\lambda \left( t\right) =\left( 1-t\right) D_{0}+t\rho \left( b\right) $%
, then $\lambda \left( t\right) $ is an inseparable $PPT$ state provided 
\[
t\left( b\right) \equiv s_{0}\left[ \frac{Np_{\max }-1}{N/m-1}\right] <t\leq %
1
\]
where $s_{0}=1-\frac{\epsilon N}{m}$ as above.
\end{theorem}

\textit{Proof}: The proof is again simply a matter of checking that $%
Tr\left( \mu _{0}\lambda \left( t\right) \right) <\frac{\epsilon }{m}$ when $%
t$ satisfies the given constraint, and then noticing that $\lambda \left(
t\right) $ is a convex combination of $PPT$ states. Note that $t\left(
m\right) =s_{0}$. $\;\Box $

\section{\textbf{The non-orthogonal case}}

To generalize the theory to the non-orthogonal case, we need to identify
some consequences of orthogonality in the preceding analysis. We do that in
the subsequent paragraphs, providing an analogous methodology for
constructing inseparable densities on the ``opposite'' side of $S$ from a
particular face $F$. What is lost in this generality, however, is that the
resulting inseparable densities are not necessarily $PPT$. In fact, one can
use this ``far-face'' methodology to represent the maximally entangled state
for two qubits $\left| \psi \right\rangle \left\langle \psi \right| $, where 
$\left| \psi \right\rangle =\frac{1}{\sqrt{2}}\left( \left| 00\right\rangle
+\left| 11\right\rangle \right) $, as the $\rho _{0}$ obtained from a
separable $\mu _{0}$.

We continue with the notation that $B$ denotes a set of $m$ separable
vectors $\left| \varphi _{k}\right\rangle $ but no longer require that they
be orthogonal. However, we continue to assume that $B$ is unextendible.

\textit{Condition} \textit{1}: $B^{\bot }$ contains no separable vectors.

One would think that reducing the restrictions on states in $B$ would make
it easier to find examples, and that seems to be the case. Rather than
working in maximum generality, however, we restrict our attention to $%
H^{\left[ N\right] }=H^{\left[ d\right] }\otimes H^{\left[ d\right] }$ and
record a result found in \cite{BD1}.

\begin{lemma}
Let $B=\left\{ \left| \varphi _{n}\right\rangle =\left| \alpha
_{n}\right\rangle \otimes \left| \beta _{n}\right\rangle ,1\leq n\leq %
2d-1\right\} $ satisfy the following property: 
\begin{eqnarray}
&&\text{every subset of size }d\text{ of }\left\{ \left| \alpha
_{n}\right\rangle ,1\leq n\leq 2d-1\right\}   \label{extendcond} \\
&&\text{and of }\left\{ \left| \beta _{n}\right\rangle ,1\leq n\leq %
2d-1\right\} \text{ is a basis for }H^{\left[ d\right] }  \nonumber
\end{eqnarray}
Then there is no separable projection $\left| \varphi \right\rangle =\left|
\alpha \right\rangle \otimes \left| \beta \right\rangle $ in $B^{\bot }$.
\end{lemma}

\textit{Proof}: If $\left\langle \varphi \right. \left| \varphi
_{n}\right\rangle =0$, for each $n$, then there is a subset of indices of
size $d$ such that either $\left\langle \alpha \right. \left| \alpha
_{n}\right\rangle =0$ for all such $n$ or $\left\langle \beta \right. \left|
\beta _{n}\right\rangle =0$ for all such \textit{n}. But any vector
orthogonal to a basis is necessarily zero, proving the point. \ $\Box $

Another consequence of the orthogonality assumption is that $\mu _{0}$ is a
density and is in $F\left( B\right) $. A weaker condition gives the same
result, and we should point out that it may not even be necessary in the
analysis to require that $\mu _{0}$ is actually in $F\left( B\right) $.

\textit{Condition} \textit{2}: There exists an $m$-vector $p$ with
non-negative entries such that $\sum_{k}p_{k}=1$ and $\sum_{k=1}^{m}\left|
\left\langle \varphi _{r}\right| \varphi _{k}\rangle \right| ^{2}p_{k}$ is
constant.

There are equivalent versions of this condition which may make the
motivation clearer. One version is that there is a density 
\begin{equation}
\mu _{0}=\sum_{k}p_{k}\mu _{k}  \label{mu0}
\end{equation}
such that $Tr\left( \mu _{0}^{2}\right) =Tr\left( \mu _{r}\mu _{0}\right) $
for all $r$. Another version is that the positive convex cone defined by the
columns of the quadratic form $Q\left( r,k\right) =Tr\left( \mu _{r}\mu
_{k}\right) =\left| \left\langle \varphi _{r}\right| \varphi _{k}\rangle
\right| ^{2}$ contains a constant vector. In the case when the vectors $%
\left| \varphi _{k}\right\rangle $ are orthogonal, these conditions are
easily satisfied, and there is the same geometric interpretation in the
non-orthogonal case.

\begin{lemma}
\textit{Condition} \textit{2} is equivalent to the property that the ``line
segment'' from $\mu _{0}$ through $D_{0}$ is orthogonal to $F\left( B\right) 
$.\ $\Box $
\end{lemma}

This condition is also relatively easy to satisfy, and the basic requirement
is that the values of $\left| \left\langle \varphi _{r}\right| \varphi
_{k}\rangle \right| ^{2}$ aren't too large.

\begin{lemma}
Suppose that $\sum_{k\neq r}\left| \left\langle \varphi _{r}\right| \varphi
_{k}\rangle \right| ^{2}\leq t<1$ for all values of $r$. Then there is a
strictly positive probability vector $p$ satisfying \textit{Condition} 
\textit{2}.
\end{lemma}

\textit{Proof}: With $Q\left( r,k\right) =\left| \left\langle \varphi
_{r}\right| \varphi _{k}\rangle \right| ^{2}$, let $B=Q-I$, where $I$ is the
identity and thus $B$ is non-negative and zero down the main diagonal. It
follows from $\sum_{k=1}^{m}B\left( r,k\right) \leq t$ and an induction
argument that $\sum_{k=1}^{m}B^{\left( n\right) }\left( r,k\right) \leq
t^{n} $ for the iterates of $B$. Let $e$ denote the vector with coordinates
equal to $1$. Then the equation $Qx=e$ has the solution 
\[
x=\left( I+B\right) ^{-1}e=\sum \left( -1\right) ^{k}B^{k}e=\sum
B^{2k}\left( e-Be\right) . 
\]
Since $e-Be$ is strictly positive, so is $x$, and $p=x/\sum x_{k}$ is the
desired probability vector.\ $\Box $

\begin{corollary}
Under the same hypothesis, $F\left( B\right) $ is a simplex: each $\mu $ in $%
F\left( B\right) $ has a unique convex representation in terms of the $\mu
_{k}$'s.
\end{corollary}

\textit{Proof}: If $\mu =\sum_{k}p_{k}\mu _{k}=\sum_{k}q_{k}\mu _{k},$ then
for all $j$%
\[
Tr\left( \mu \mu _{j}\right) =\sum_{k}Q\left( j,k\right)
p_{k}=\sum_{k}Q\left( j,k\right) q_{k}. 
\]
Since $Q$ is invertible, the assertion is immediate. \ $\Box $

Combining the first two conditions gives the analogue of Proposition 3.1.
However, since the spectral representation of $\mu _{0}$ no longer coincides
with its convex representation, we need to introduce explicitly the
eigenvalues $\lambda _{k}$ of $\mu _{0}$ with $\lambda _{max}$ denoting the
largest eigenvalue. With exactly the same proof as before, we then have the
following result.

\begin{proposition}
inf$\left\{ Tr\left( \mu _{0}\sigma \right) ,\sigma \in S\right\} \equiv
\epsilon \lambda _{max}>0$, and the non-empty compact convex subset of $S$ 
\[
G\left( B\right) =\left\{ \sigma \in S:Tr\left( \mu _{0}\sigma \right)
=\epsilon \lambda _{max}\right\} 
\]
is contained in an affine set orthogonal to the line from $\mu _{0}$ through 
$D_{0}$.\medskip \ $\Box $
\end{proposition}

Define $b=1/\lambda _{max}$, so that $b\leq m$, and set 
\[
\rho _{0}=\frac{1}{N-b}\left( ND_{0}-b\mu _{0}\right) 
\]
as before. Using the spectral representation of $\mu _{0}$, which is now
distinct from its convex representation, familiar arguments confirm the
following result. Note that we do not assert that $\rho _{0}$ is $PPT$ or
even inseparable.

\begin{lemma}
$\rho _{0}$ is a density on the boundary of $D$.\ $\Box $
\end{lemma}

\textit{Conditions} \textit{1} and \textit{2} are easily satisfied, but
dropping orthogonality introduces a third requirement which is much more
restrictive, and this final condition is necessary to complete the extension
to the non-orthogonal $UPB$ case. The condition depends heavily on the
eigenvalues of $\mu _{0}$, a fact that is not immediately obvious in the
proof of the orthogonal case and which is necessary to obtain the analogue
of (\ref{epsineq}). In the orthogonal case, the right hand side below is
zero, and the inequality follows from $\epsilon >0.$

\textit{Condition} \textit{3}: $\epsilon \lambda _{max}>\left( \lambda
_{max}-Tr\left( \mu _{0}^{2}\right) \right) /\left( N\lambda _{max}-1\right) 
$.

The reasoning behind (\ref{neg}) still applies and this time gives 
\begin{equation}
0<\epsilon N\lambda _{max}<1,  \label{epslamda}
\end{equation}
setting the stage for the final bit of analysis.

\begin{theorem}
Suppose the set of separable states $B$ satisfies \textit{Conditions 1, 2, }%
and\textit{\ 3}. Let 
\[
s_{0}=\frac{\left( 1-\epsilon N\lambda _{max}\right) \left( N\lambda
_{max}-1\right) }{NTr\left( \mu _{0}^{2}\right) -1}.
\]
Then $0<s_{0}<1$. Define $\tau _{0}=\tau _{0}\left( s_{0}\right) =\left(
1-s_{0}\right) D_{0}+s_{0}\rho _{0}$ and use the usual notation to define $%
W_{0}=\tau _{0}+c_{0}I-\rho _{0}$. Then $\tau _{0}$ is a density, and $W_{0}$
is an entanglement witness for $\rho _{0}$, which is therefore inseparable.
\end{theorem}

\textit{Proof}: Each of the factors defining $s_{0}$ is positive, so we only
need check that $s_{0}<1$. Working out the algebra, which we omit, shows
that $s_{0}<1$ is equivalent to \textit{Condition} \textit{3}, and thus we
know that $\tau _{0}$ lies strictly between $D_{0}$ and $\rho _{0}$,
although we cannot claim that $\tau _{0}$ is itself separable. Once we
verify that $Tr\left( \mu _{0}\tau _{0}\right) =\epsilon \lambda _{max}$,
which is a straight-forward calculation, the logic follows the pattern of
the analogous result in the orthogonal case, completing the proof.\ $\Box
\smallskip $

I am indebted to the referee for correcting several misstatements in an
earlier version of this paper and also for asking for examples illustrating
the theory of this section. This led to the results above which show that it
is quite easy to give examples of sets $B$ satisfying the first two
conditions. In fact, we give an example of a $B$ in the $2\otimes 2$ case
which satisfies \textit{Conditions} \textit{1} and \textit{2}, something
that is not possible when orthogonality is required (\cite{BD1}).

\begin{example}
Let $d=2$ and define the three states $\left| \varphi _{n}\right\rangle $, $1%
\leq n\leq 3$, by $\left| \alpha _{1}\right\rangle =\frac{1}{\sqrt{2}}{%
1 choose 1}=\left| \beta _{1}\right\rangle $, $\left| \alpha _{2}\right\rangle =%
\frac{1}{\sqrt{2}}{1\choose -1}=\left| \beta _{2}\right\rangle $, $\left|
\alpha _{3}\right\rangle =\frac{1}{\sqrt{2}}{1 \choose i}$, and $\left| \beta
_{3}\right\rangle \frac{1}{\sqrt{2}}{1 \choose -i}$. Then $B_{2}$ satisfies (%
\ref{extendcond}), the associated $Q$ matrix is $\left( 
\begin{array}{ccc}
1 & 0 & 1/4 \\ 
0 & 1 & 1/4 \\ 
1/4 & 1/4 & 1
\end{array}
\right) ,$ and the $p$-vector is $\left( 3/8,3/8,2/8\right) $. \ $\Box $
\end{example}

The real difficulty is with \textit{Condition} 3, and there is no guarantee
that a $B$ satisfying the first two conditions will also satisfy the third.
In fact one can show that \textit{Condition} \textit{3} does not hold in the
example above. To illustrate a methodology which simplifies the calculation
of $\lambda _{max}$, we provide the details.

\begin{lemma}
Suppose \textit{Conditions} \textit{1} and \textit{2} are satisfied and the $%
p$-vector is strictly positive. Then the positive eigenvalues of $\mu _{0}$
coincide with the positive eigenvalues of $R$ where 
\[
R\left( r,n\right) =p_{r}\left\langle \varphi _{r}\right| \varphi
_{n}\rangle .
\]
\end{lemma}

\textit{Proof}: If $\mu _{0}\left| \psi \right\rangle =\lambda \left| \psi
\right\rangle $ for positive $\lambda $, then necessarily $\left| \psi
\right\rangle $ is in the span of the $\left| \varphi _{n}\right\rangle $'s: 
$\left| \psi \right\rangle =\sum x_{n}\left| \varphi _{n}\right\rangle $.
Rewriting the eigenvalue equation we obtain 
\[
\sum_{r}\left| \varphi _{r}\right\rangle \left[ \sum_{n}R\left( r,n\right)
x_{n}-\lambda x_{r}\right] =0. 
\]
Since $Tr\left( R\right) =1$, if $R$ has non-negative eigenvalues, then its
positive eigenvalues necessarily coincide with those of $\mu _{0}$. Using
the strict positivity of the components of the probability vector $p$, $R=D%
\tilde{R}D^{-1}$ where $\tilde{R}\left( r,n\right) =\sqrt{p_{r}}\left\langle
\varphi _{r}\right| \varphi _{n}\rangle \sqrt{p_{n}}$ and the diagonal
matrix $D$ has entries $\sqrt{p_{r}}$. But $\tilde{R}$ is a trace one
positive semi-definite matrix whose eigenvalues coincide with those of $R$,
and that completes the proof. Note that this approach does not require that
the $\left| \varphi _{n}\right\rangle $ be linearly independent. \ $\Box $

\begin{example}
In the example from above, one has 
\[
R=\left( 
\begin{array}{ccc}
3/8 & 0 & 3/16 \\ 
0 & 3/8 & 3/16 \\ 
1/8 & 1/8 & 1/4
\end{array}
\right) 
\]
and computes that $\mu _{0}$ has positive eigenvalues $\left( 5\pm \sqrt{13}%
\right) /16$ and $3/8$. The right-hand side of the inequality in \textit{%
Condition} \textit{3} equals $\left( 5-\sqrt{13}\right) /16$ and the infimum
of $Tr\left( \mu _{0}\sigma \right) $ appears to be $1/16$, when $\sigma $
is the density associated with $\left| \alpha _{1}\right\rangle \otimes
\left| \beta _{2}\right\rangle $. In any event, \textit{Condition} \textit{3}
does not hold, and the associated $\rho _{0}$ is on the same side of the $%
W_{0}$ hyperplane as $\mu _{0}$. In fact, one can show that $\rho _{0}$ is
separable. \ $\Box $
\end{example}

To get a positive result, we can perturb examples from the orthogonal case.
The idea is to take an orthogonal \textit{UPB} $B$ and slightly modify some
of the components of the $\left| \varphi _{n}\right\rangle $'s using a
parameter $t$ so that the unextendibility is not lost. If this is done so
that $\mu _{0}\left( t\right) $ and its eigenvalues converge to those in the
original set $B$ as $t$ goes to $0$, then \textit{Condition} \textit{3} will
be satisfied provided \textit{t} is small enough: 
\begin{eqnarray*}
Tr\left( \mu _{0}\left( t\right) \sigma \right) &=&Tr\left( \mu _{0}\sigma
\right) +Tr\left( \left( \mu \left( t\right) -\mu _{0}\right) \sigma \right)
\\
&\geq &\epsilon \lambda _{max}+Tr\left( \left( \mu \left( t\right) -\mu
_{0}\right) \sigma \right) \\
&\gtrsim &\left( \lambda _{max}\left( t\right) -Tr\left( \mu _{0}^{2}\left(
t\right) \right) \right) /\left( N\lambda _{max}\left( t\right) -1\right)
\rightarrow 0.
\end{eqnarray*}

\begin{example}
Take for $B$ the orthogonal ``TILES'' of the $3\times 3$ case in \cite{BD1}: 
$\left| \varphi _{1}\right\rangle =\frac{1}{\sqrt{2}}\left| 0\right\rangle
\left( \left| 0\right\rangle -\left| 1\right\rangle \right) $, $\left|
\varphi _{2}\right\rangle =\frac{1}{\sqrt{2}}\left| 2\right\rangle \left(
\left| 1\right\rangle -\left| 2\right\rangle \right) $, $\left| \varphi
_{3}\right\rangle =\frac{1}{\sqrt{2}}\left( \left| 0\right\rangle -\left|
1\right\rangle \right) \left| 2\right\rangle $, $\left| \varphi
_{4}\right\rangle =\frac{1}{\sqrt{2}}\left( \left| 1\right\rangle -\left|
2\right\rangle \right) \left| 0\right\rangle $, and $\left| \varphi
_{5}\right\rangle =\left| \gamma \right\rangle \left| \gamma \right\rangle $
where $\left| \gamma \right\rangle =\frac{1}{\sqrt{3}}\left( \left|
0\right\rangle +\left| 1\right\rangle +\left| 2\right\rangle \right) $.
Modify $\left| \varphi _{5}\right\rangle $ by setting $\left| \varphi
_{5}\right\rangle \left( t\right) =\left| \gamma \right\rangle \frac{1}{%
\sqrt{c\left( t\right) }}\left( \left( 1+t\right) \left| 0\right\rangle
+\left| 1\right\rangle +\left| 2\right\rangle \right) $ where $c\left(
t\right) $ is the appropriate normalizing factor. Straightforward
computations give 
\[
Q=\left( 
\begin{array}{ccccc}
1 & 0 & 0 & 0 & \frac{t^{2}}{6c\left( t\right) } \\ 
0 & 1 & 0 & 0 & 0 \\ 
0 & 0 & 1 & 0 & 0 \\ 
0 & 0 & 0 & 1 & 0 \\ 
\frac{t^{2}}{6c\left( t\right) } & 0 & 0 & 0 & 1
\end{array}
\right) \text{and }p=\frac{1}{5+\frac{t^{2}}{2c\left( t\right) }}\left( 
\begin{array}{c}
1 \\ 
1+\frac{t^{2}}{6c\left( t\right) } \\ 
1+\frac{t^{2}}{6c\left( t\right) } \\ 
1+\frac{t^{2}}{6c\left( t\right) } \\ 
1
\end{array}
\right) 
\]
The eigenvalues are easily computable using the $R$ matrix and are
continuous functions of $t$ which converge to $1/5$. Moreover, 
\[
\left( \lambda _{max}\left( t\right) -Tr\left( \mu _{0}^{2}\left( t\right)
\right) \right) /\left( 9\lambda _{max}\left( t\right) -1\right) =\frac{t}{4%
\sqrt{6c\left( t\right) }}r\left( t\right) 
\]
where $r\left( t\right) $ is a rational function converging to $1$ as $%
t\rightarrow 0$. Thus, for sufficiently small $t$, which depends on the
value of $\epsilon $, \textit{Condition} \textit{3} is satisfied.
\end{example}

\textit{Acknowledgments}: I am indebted to M. Rubin for useful discussions
and for pointing out the role of the ``far face'' of $S$ in the analysis of
inseparable densities and to S. Gowda for a delightful discussion which led
to the proof of Lemma 4.3. Much of the research for this paper was completed
during a visit in the summer of 2001 to the Oxford Centre for Quantum
Computation, and the Centre's hospitality is gratefully acknowledged. In
independent work, the role of \textit{UPB} bases in constructing \textit{PPT}
densities has also been investigated recently by S. Bandyopadhyay, S. Ghosh,
and Y. P. Rowchowdhury at UCLA.

\end{document}